\documentclass[osajnl,twocolumn,showpacs,superscriptaddress,11pt]{revtex4-1} 
\usepackage{amsmath,amssymb,graphicx}

\begin{document}

\title{Transmission and reflection characteristics of metal-coated optical fiber tip pairs}

\author{Jean-Baptiste Decombe}
\author{Jean-Fran\c{c}ois Bryche}
\author{Jean-Fran\c{c}ois Motte}
\author{Jo\"{e}l~Chevrier}
\author{Serge~Huant}
\author{Jochen Fick}\email{Corresponding author: jochen.fick@grenoble.cnrs.fr}
\affiliation{ Institut N\'{e}el, CNRS \& Universit\'{e} Joseph Fourier, 25 Avenue des Martyrs, 38042~Grenoble, France }

\begin{abstract}
	The optical transmission and reflection in between two metalized optical fiber tips is studied in the optical near-field and far-field domains. Besides aluminum-coated tips for near-field scanning optical microscopy (NSOM), specifically developed gold-coated fiber tips cut by focused ion beam (FIB) are investigated. Transverse transmission maps of sub-wavelength width clearly evidence optical near-field coupling between the tips for short tip distances and becomes essentially Gaussian-shaped for larger distances in the far-field regime. Moreover concentric reflection fringes observed for NSOM-type tips illustrate the influence of the receiving fiber tip on the emission pattern of the source tip.
\end{abstract}


\maketitle 

\section{Introduction}

Optical fiber conical  tips with a sub-wavelength clear aperture at the apex are common tools in micro- and nano-optics. They are used for optical trapping of micro-particles in single-fiber-tip \cite{SSP+12,ETH09,LGY+06} and counter-propagating two-fiber-tip \cite{VOO09,LS95} configurations. Fiber tips are also used in scanning optical microscopy: bare fiber tips can be applied for imaging of neurons \cite{DSV+11}, whereas metalized tips are the key element of NSOM \cite{BTH+91,OK95}.

The optical near field of metalized fiber tips was probed using fluorescent nanospheres and an analytical model for the emitted electric field was developed \cite{DNH+04}. The near-field intensity shows two lobes whereas the far-field emission is polarization dependent and of excellent Gaussian shape with large emission angles exceeding 90$^\circ$ in the P polarization \cite{OK95,DWH02}. The emission angle can be directly linked to the tip apex size. These features have been essentially recovered by FDTD calculations \cite{AS07}. To some extent, a sub-wavelength aperture at the apex of a conical tip mimics a single small diffraction hole on a flat metallic screen, a problem that has been extensively studied \cite{YCdL+12, KKK+11} since the pioneer work of Bethe \cite{Bet44}.

Recently the power propagation in apertureless metal-coated optical
fiber tips was investigated theoretically \cite{BFB+12} and tip-to-tip scans of such fiber tips were studied in view of their applications to NSOM lithography \cite{KPS+13}.

Two open aperture metalized fiber tips facing each other at distance of some hundred nanometers are a promising approach for optical nano-tweezers. They combine the advantages of nano-traps based on a plasmonic cavity \cite {PG12,TKS13} with the flexibility of fiber based optical tweezers \cite{LGY+06}. This makes it possible to realize a genuine plasmonic tweezers, allowing not only nanoparticle trapping but also their manipulation at the nanoscale, with the rewarding  prospect of possible operation in air, not only in a liquid.

In this paper we present the optical transmission and reflection study of metalized fiber tip pairs with tip distances in between tens of nanometers up to tens of microns, thus covering the optical near field and far field ranges. More specifically the transition between these two regimes is studied. Such tip pairs are aimed at being used in future near-field optical tweezers.

\section{Experimental}

Two different types of metalized fiber tips are studied: NSOM tips and FIB-cut tips. The first one consists of open aperture, aluminum-coated fiber tips currently used for NSOM. The second one are gold-coated fiber tips, where the aperture is obtained by FIB-cutting of the entirely metalized fiber tips. The main difference of these two tip types consists in the shape of the tip apex. The application of two different metals does not influence the results presented in this paper.

\begin{figure}[htb]
	\centering
	\includegraphics[width=5.cm]{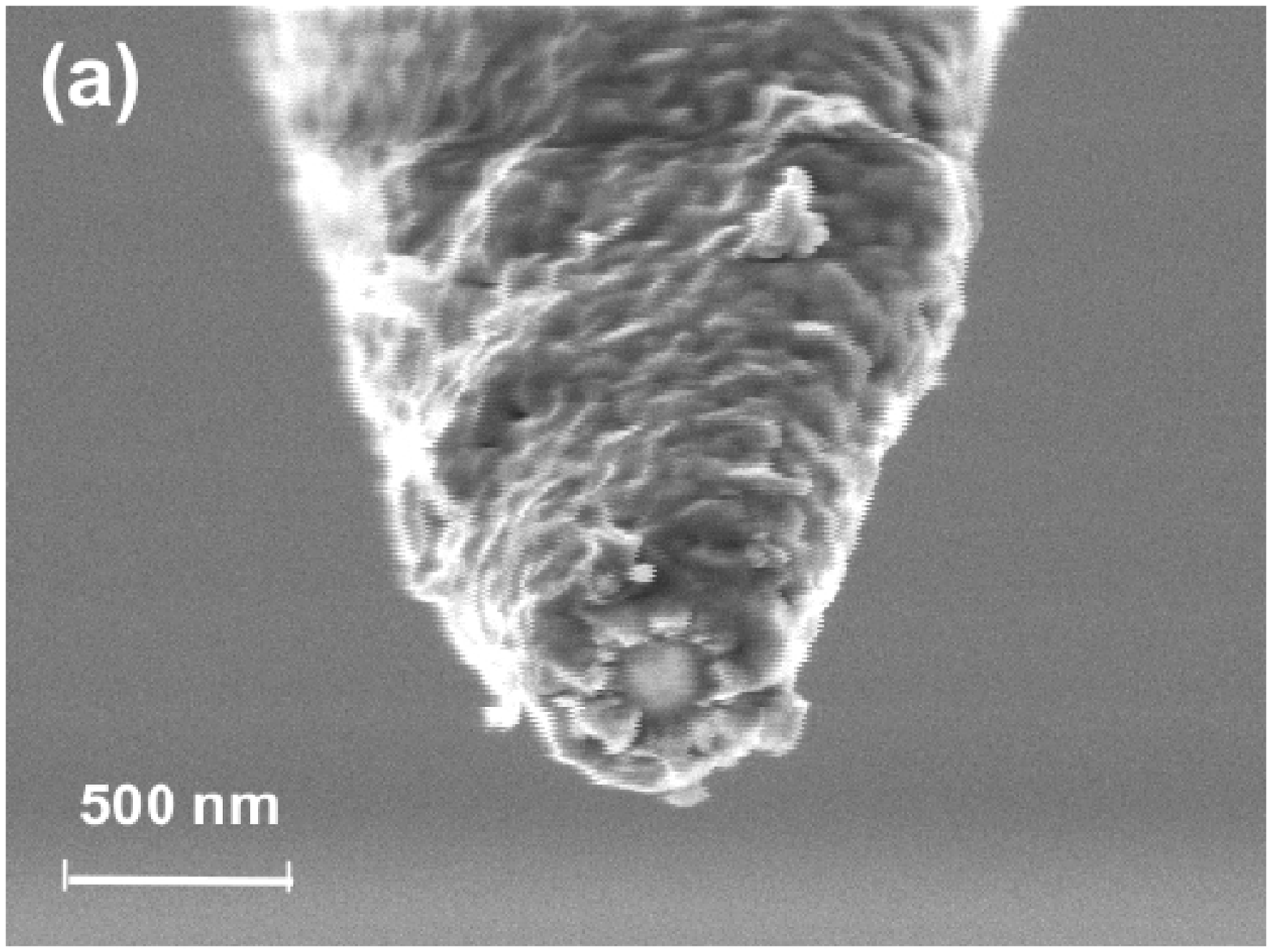}\\
	\includegraphics[width=5.cm]{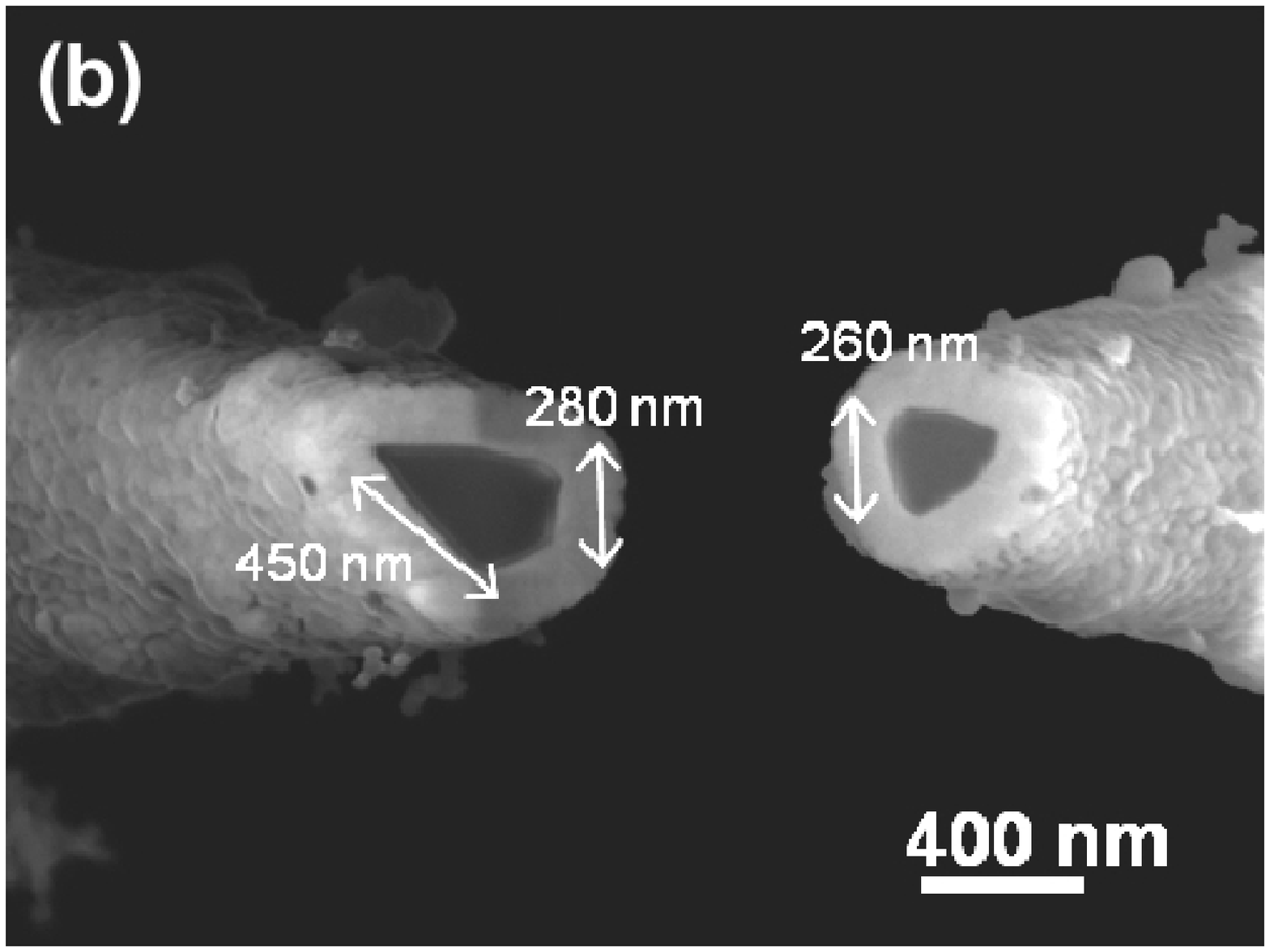}
	\caption{Scanning electron micrographs of the two fiber tip types used: (a) typical NSOM tip, and (b) the FIB-cut tips used in the experiments.\label{fig.SEM}}
	
\end{figure}

The elaboration of both tip types is based on chemical wet-etching of single mode, pure silica core fibers (Nufern S630-HP) \cite{CSM+06}. The obtained fiber tip angle is about $15^{\circ}$. In the case of NSOM tips, a magnesium fluoride (MgF$_2$) film is first deposited on the as-etched tips in order to control the final apex diameter. Then an opaque aluminum film of about 100 nm thickness is deposited by thermal evaporation after the deposition of a thin nickel-chromium adhesion layer. This technique allows to obtain apex diameters in the order of 200 - 300 nm without any further step such as FIB-cutting (Fig. \ref{fig.SEM}.a).

Cutting the fiber tips using FIB leads to smooth, high quality end-faces (Fig. \ref{fig.SEM}.b). This point is of paramount importance for experiments with tip distances in the nanometer regime, thus justifying the more complex fabrication process compared to the NSOM-tips. At the same time we start using gold for its capacity to support low loss surface plasmons. Aiming smallest achievable aperture sizes, we dismiss the thick MgF$_2$ layer. The 200 nm gold layer is deposited directly on the etched fiber tips, only using a 10 nm titanium adhesion layer. The fiber tip end is then cut by FIB to get sub-micrometer tip apertures. With this technique the obtained tips are of elliptical or nearly circular shape, thus allowing to study shape effects.

\begin{figure}[htb]
	\centering
	\includegraphics[width=8.cm]{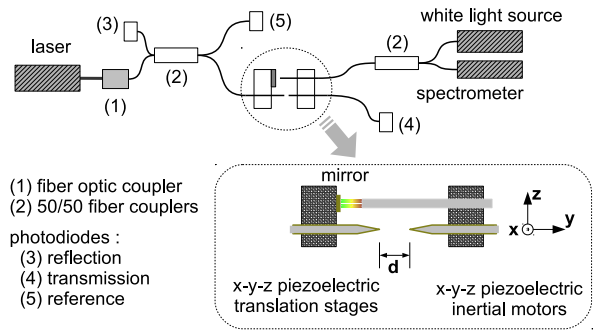}\\

	\caption{Scheme of the experimental set-up.}
	\label{fig.setup}
\end{figure}

The transmission and reflection of optical fibers are measured on a dedicated set-up allowing scanning the relative fiber position with nanometer accuracy (Fig. \ref{fig.setup}). One fiber is mounted on a set of $xyz$ piezoelectric translation stages (PI P620) with sub-nanometer resolution and 50 $\mu$m range. The second fiber is mounted on three perpendicular inertial piezoelectric translation stages (Mechonics MS~30) with $\approx 30$ nm step size and up to 2.5 cm range. A microscope with a long working distance objective (Mitutoyo M 50x) coupled to a CMOS camera allows to visualize the fiber tips with micrometer resolution.

For technical reasons we use two different diode lasers emitting at 808 nm and 830 nm to characterize the Al- and Au-coated fiber tips, respectively. The small wavelength difference has no influence on the results as the wavelengths are far away from metal absorption bands or fiber cut-off wavelengths. A 50/50 waveguide coupler is used to allow simultaneous reflection and transmission measurements by means of two amplified Si-photodiodes (New Focus 2001-FC).

A closed-loop control is implemented to stabilize the relative fiber tip distance at the nanometer scale. In fact, thermal drifts are a serious issue in a room temperature uncontrolled environment. The axial fiber tip distance is the most critical one as the transverse position can be calibrated by using the transmission maximum position. The feedback signal is obtained from a Fabry-Perrot cavity built by a cleaved optical fiber and a metallic mirror, respectively mounted on the two fiber tip holders (Fig. \ref{fig.setup}). A fiber-coupled white lamp source and a mini-spectrometer (Avantes) allow to measure the reflection spectra of the Fabry-Perrot cavity. Its Fourier transform directly gives the absolute cavity size.

Using the closed-loop control, the relative fiber tip distance can be controlled  with a precision better than $\pm$ 5 nm over more than five hours. The absolute tip distance is, however, more difficult to assess. The only accurate way is to perform transverse scans with decreasing distance until the tips are touching. This contact can be clearly observed on the microscope image or by the appearance of streaks in the transmission intensity plot. However, the fiber tip fragility may result in severe damage. After observation of a number of contact events, we estimate that the absolute distance can be determined from the microscope images with 50-100 nm precision. As a consequence, the minimal distance of the transverse scans used in Section \ref{chap_rd} to determinate $w_0$ is of the same range.

\begin{figure}[htp]
	\centering
	\includegraphics[width=3.5cm,height=3.5cm]{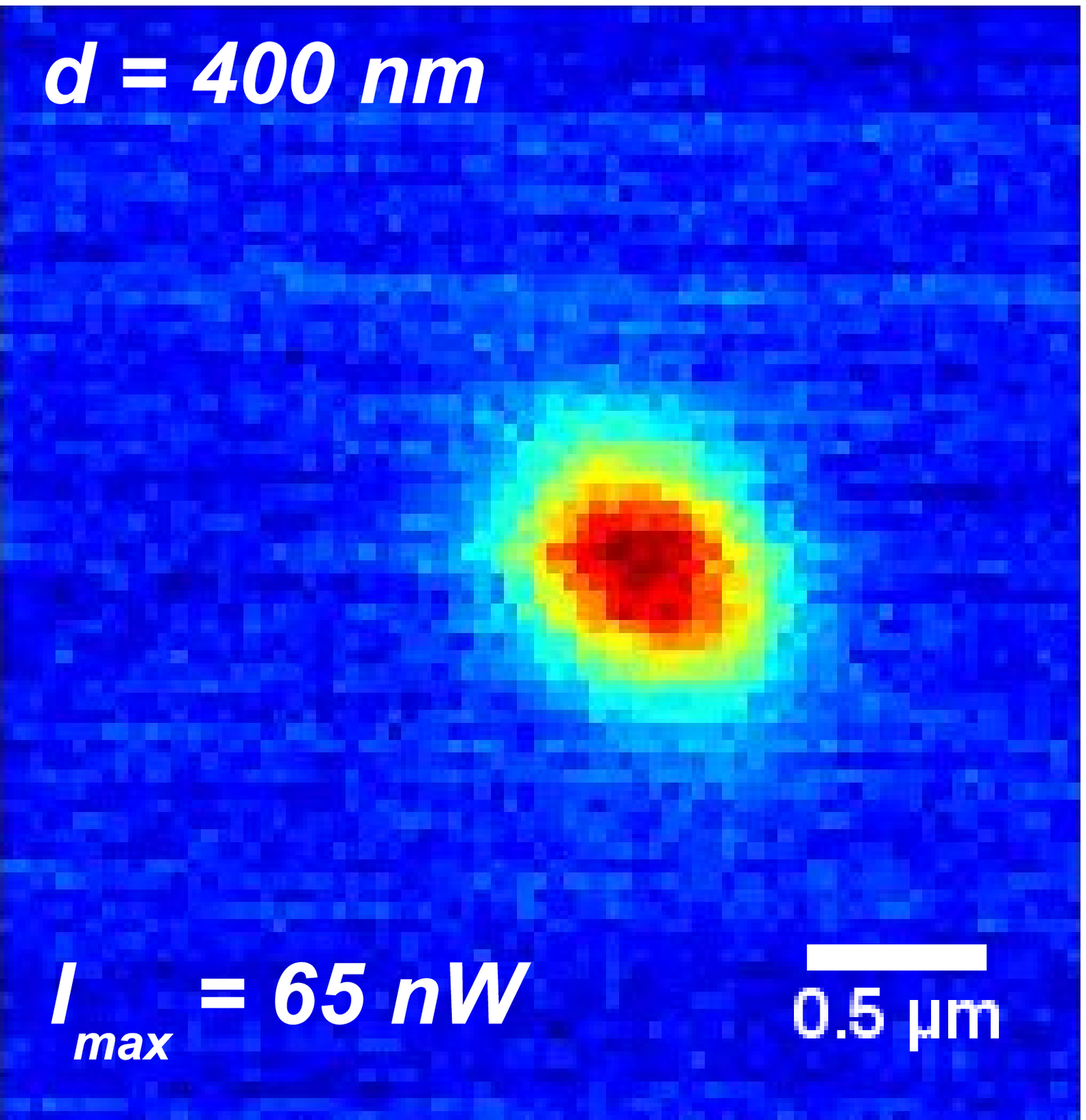}
	\includegraphics[width=3.5cm,height=3.5cm]{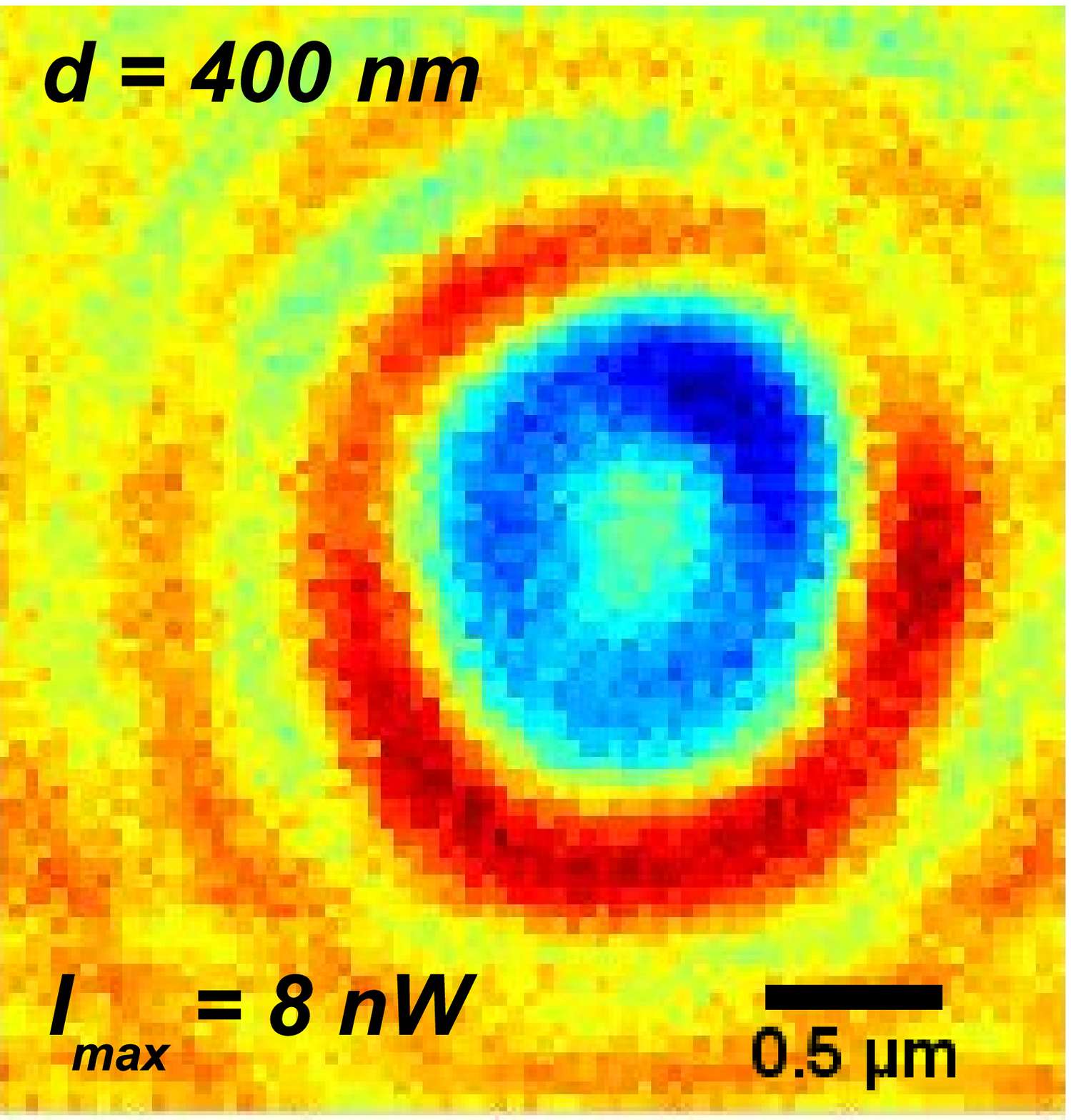}\\
	\includegraphics[width=3.5cm,height=3.5cm]{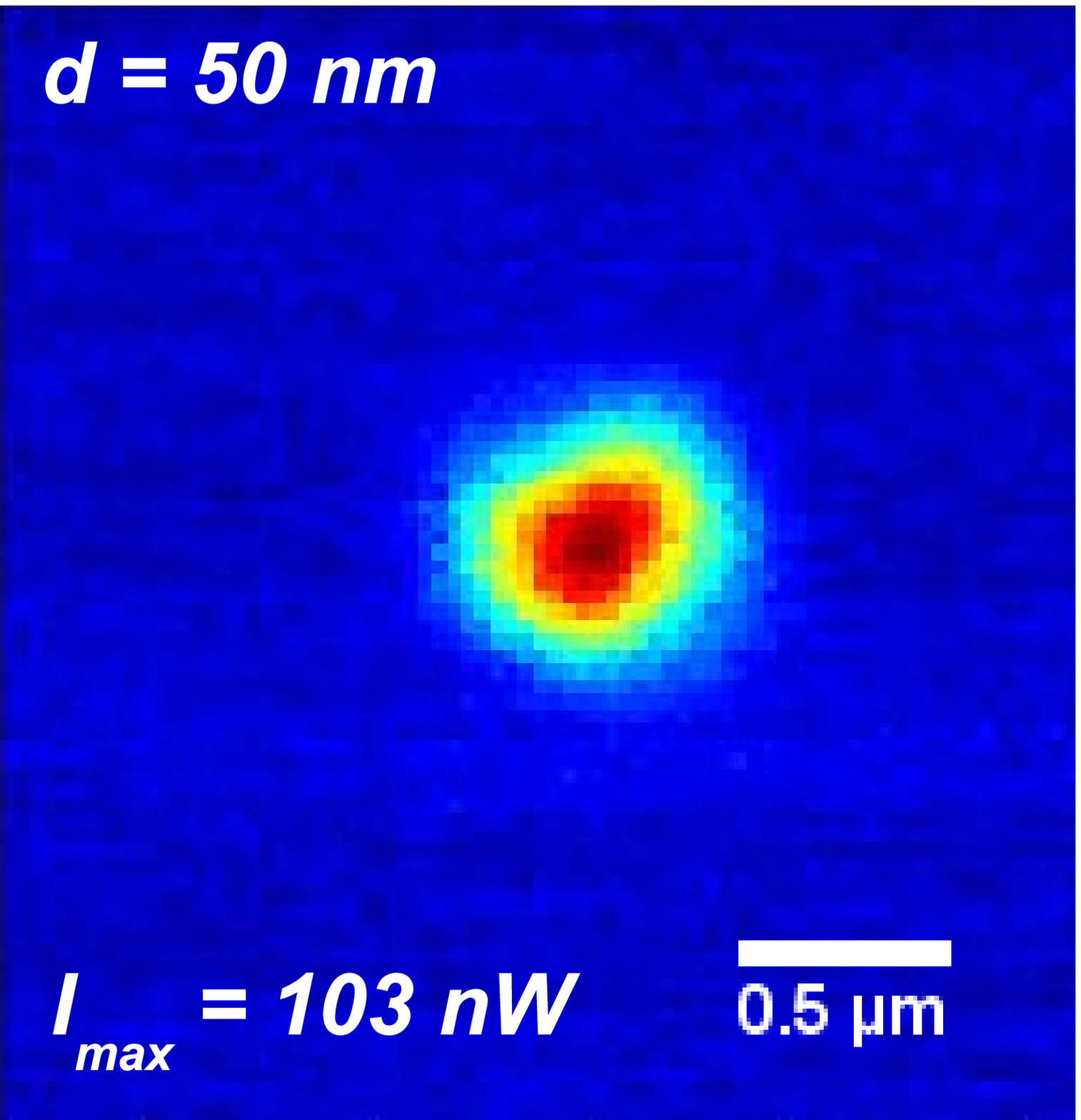}
	\includegraphics[width=3.5cm,height=3.5cm]{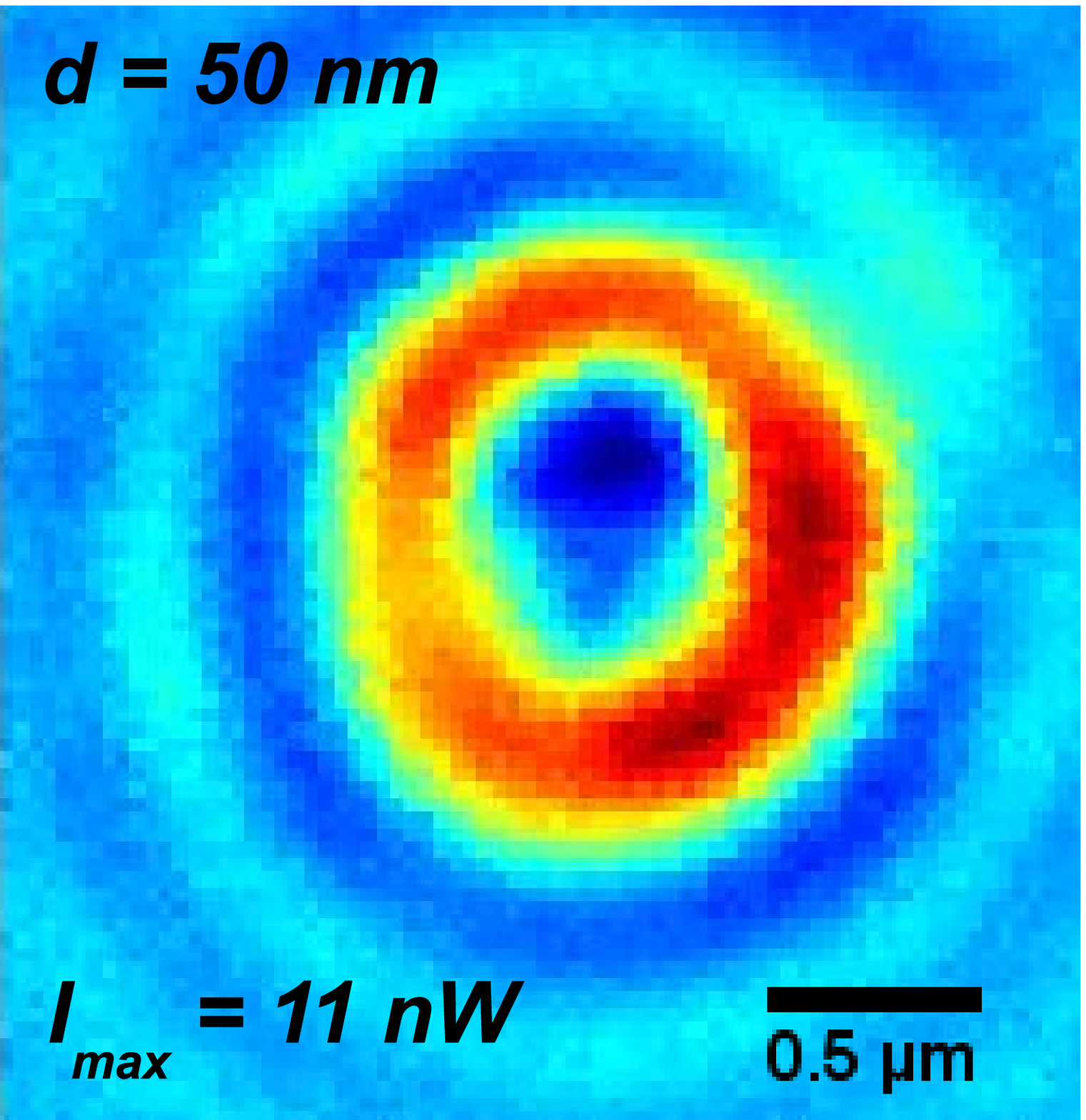}
	\caption{Transverse transmission (left) and reflection (right) intensity maps of a NSOM tip pair at two distances $d$. The maximal intensities are indicated for an injected power of P$_{in}= 300$ $\mu$W.}
	\label{fig.map}
\end{figure}

 Control software in the LabView environment allows controlling the entire set-up and recording the intensity maps. Transverse transmission and reflection maps for constant distance $d$ are recorded  by scanning one fiber in a plane perpendicular to the fiber tips' orientation. Typical scans contain $75\times 75$ data points. The intensity is averaged over 500 points with  read-out frequency of 10 kHz and a photodiode internal amplification of $50-70$ dB.

The obtained intensity plots are fitted to the Gaussian intensity beam profile function

\begin{equation}
	I(r)=I_0\cdot e^{\frac{-2(r-r_0)^2}{w^2}}
\end{equation}
with $I_0$ the intensity amplitude, $r_0$ the beam position, and $w$ the beam waist.

The recorded transmission maps correspond to the convolution of the emission and the capture functions of the two fiber tips. Thus, in order to deduce the emission spot width of only one fiber the measured transmission width has to be corrected. The corresponding deconvolution of two Gaussian functions can be done with :

\begin{equation}
  \tilde w=\sqrt{w^2-w_0^2}\label{eq.cor}
\end{equation}

with $\tilde w$ and $w$ the corrected and measured waists, respectively. $w_{0}$ is the optical aperture size of the second fiber tip. For two identical tips $w_{0}$ can be obtained by $w_0=w^{min}/\sqrt{2}$, $w^{min}$ being the measured spot size at smallest tip distances.

\section{Results and discussion \label{chap_rd}}
\subsection{NSOM fiber tips}

Transmission and reflection maps of NSOM fiber tips are recorded for tip distances up to 30 $\mu$m. Typical results are shown on Fig. \ref{fig.map}. The transmission spots are of slightly elliptical shape, in agreement with the imperfect circular symmetry of the fiber tips. The minimal measured waist is $w^{min}=375$ nm corresponding to a corrected waist of $\tilde w^{min}=265$ nm, about one-third of the wavelength $\lambda/3=269$ nm. The apex size determined by scanning electron microscopy (SEM) of the two applied tips are respectively 276 nm and 246 nm. The size of the transmission spot is thus clearly determined by the tip apex and not by the actual wavelength. 

\begin{figure}[htb]
	\centering
	\includegraphics[width=7.5cm]{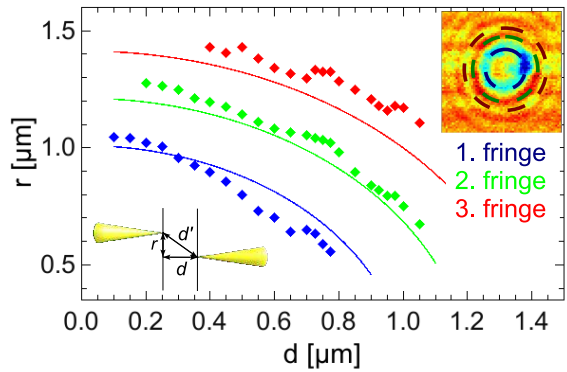}
	\caption{Position of the interference rings as a function of the axial ($d$) and transverse ($r$) fiber tips distances.}
	\label{fig.fri}
\end{figure}

The transmission spot size increases linearly with the tip distance for distances larger than $\approx 1~\mu$m. The corresponding emission angle is $\theta=18.5^\circ$, a value found independently of the fiber tip pairs. 

Clear circular and concentric reflection fringes are observed for tip distances up to several micrometers (Fig. \ref{fig.map}). The fringe center position coincides with that of the transmission peak maximum. As expected for interference fringes, intensity minima/ maxima are observed for :

\begin{equation}\label{eq.fri}
	\begin{array}{l}
   \displaystyle d^{min}_{m}=(\frac{m}{2}+\frac{1}{4})\cdot\lambda \\~\\
	\displaystyle d^{max}_{m}=\frac{m}{2}\cdot\lambda
\end{array}
\end{equation}

with $\lambda$ the wavelength and $m =1,2,...$ a positive integer.

The circular fringes can be explained by back-reflection at the fiber tips. The experimental radii of the fringe minima/ maxima are determined by fitting concentric circles to the transverse reflection intensity maps. These radii are plotted on Fig.\ref{fig.fri} as a function of the two fiber tip planes distance $d$. The theoretical fringe positions (lines in Fig. \ref{fig.fri}) are calculated using Eq. \ref{eq.fri} by substituting $d$ with $d'=\sqrt{d^2+r^2}$. The agreement with the experimental values is very good. The observed difference can be attributed to the real (elliptical) shape of the metalized fiber tips.

The observation of distinct reflection patterns clearly shows that the emission pattern of the source fiber tip is influenced by the receiving fiber tip. The reflected intensity is about one order of magnitude smaller than the transmitted intensity. However, no fringe is visible on the transmission maps. 

\subsection{FIB-cut fiber tips}

\begin{figure}[htb]
	\centering
	\includegraphics[width=3.5cm,height=3.5cm]{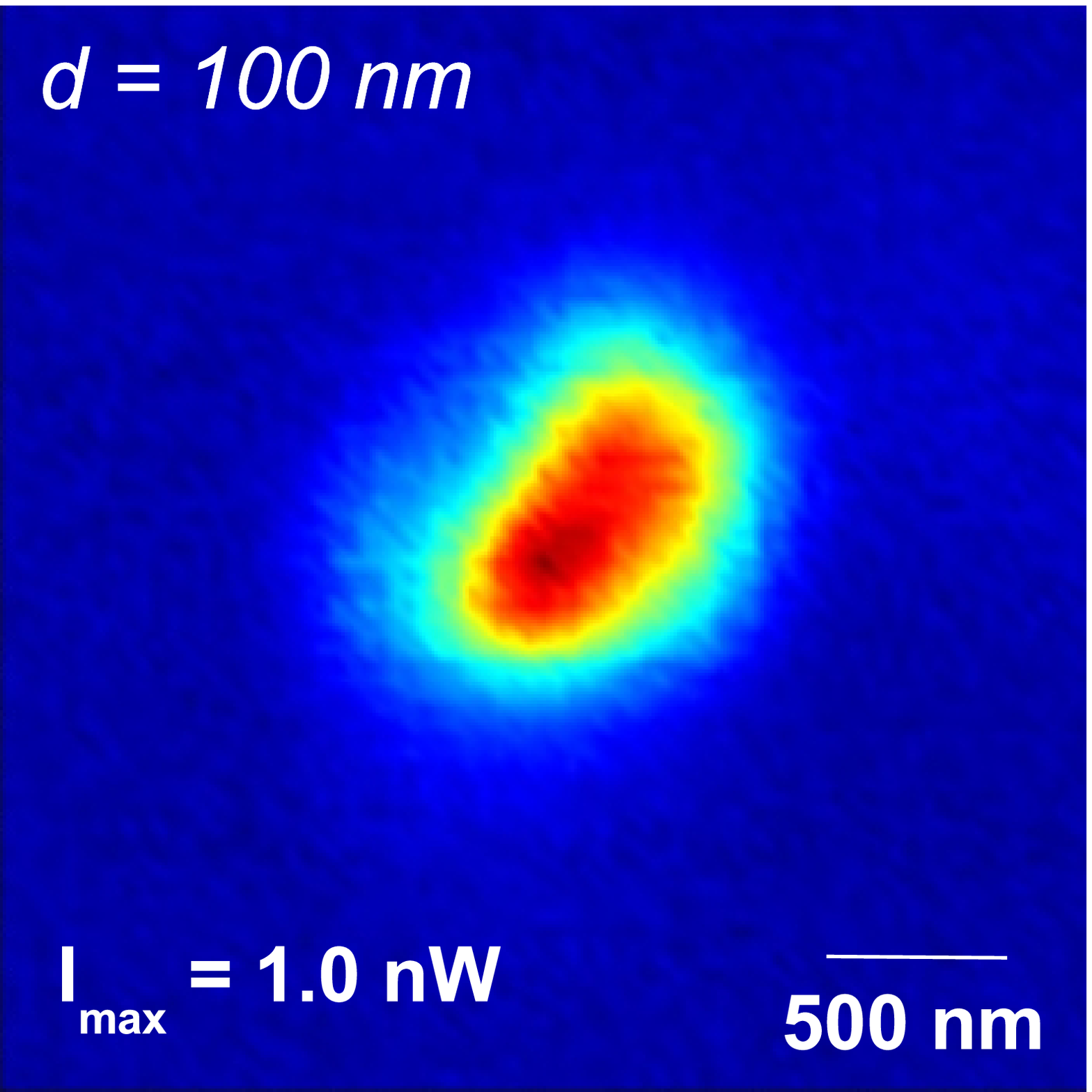}
	\includegraphics[width=3.5cm,height=3.5cm]{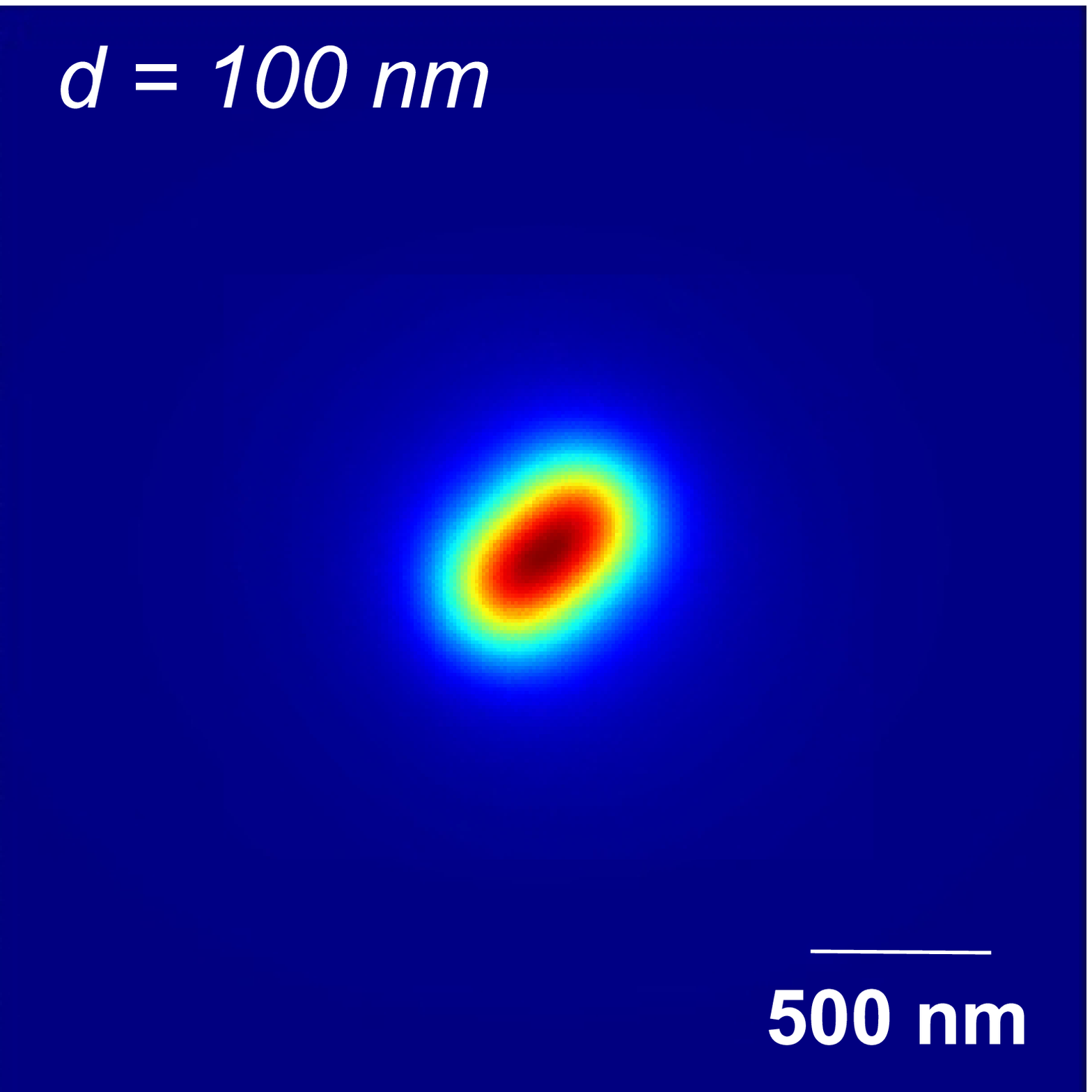}\\
	\includegraphics[width=3.5cm,height=3.5cm]{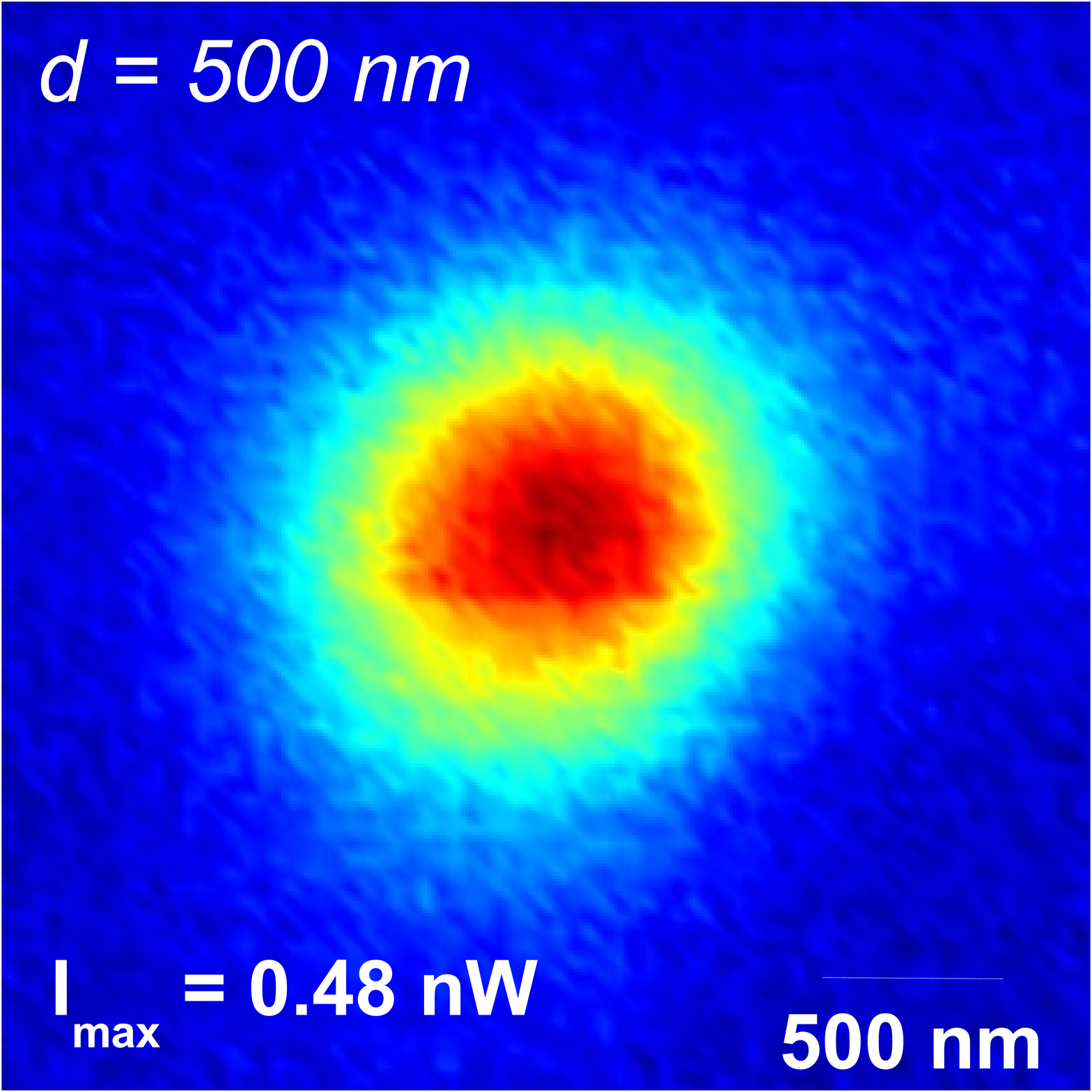}
	\includegraphics[width=3.5cm,height=3.5cm]{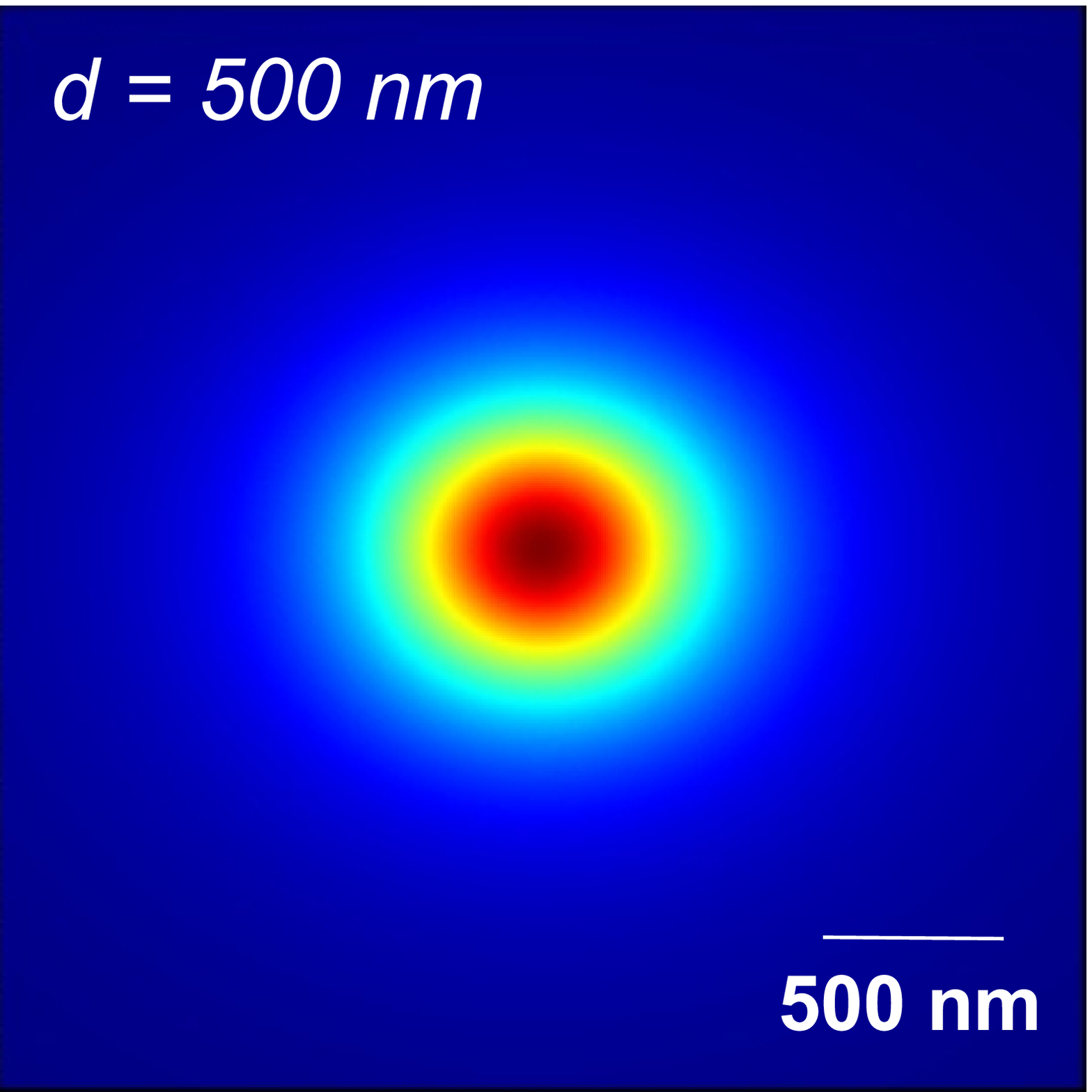}
	\caption{Experimental (left) and theoretical (right) transverse transmission intensity maps of the FIB-cut fiber tip pair shown on Fig. \ref{fig.SEM}.b. The maximal intensities are indicated for the experimental results.}
	\label{fig.au}
\end{figure}

Now, two FIB-cut fiber tips with different apex shapes are studied. The emission fiber tip is strongly elliptical with major and minor axis diameters of $a=450$ and $b=280$ nm, respectively (Fig. \ref{fig.SEM}.b, left side). The receiving tip has a nearly equilateral triangle shape with 260 nm side length . The measured transmission maps are elliptical for small tip distances and become circular for larger distances (Fig. \ref{fig.au}). The elliptical transmission spots are described by two waists ($w_a$ and $w_b$) measured parallel to the major axis $a$ and minor axis $b$, respectively. The as-measured minimal waists of the transmission spot are $w_a^{min}=470$ nm and $w_b^{min}=310$ nm.

The injected optical intensity inside the source fiber is 1 mW. The emitted intensity of the source fiber, measured by means of a power meter, is 1 $\mu$W. The transmitted maximal intensity is 1 nW and 0.48 nW for tip distances of 100 nm and 500 nm, respectively. The emission losses of 30 dB are essentially due to strong propagation losses in the sub-wavelength wide section near the fiber tip end. For the measurement at $d=100$ nm, the reception losses are of the same order than the emission losses, suggesting that the propagation losses in between the two fiber tips can be neglected. The lower maximal transmitted intensity of the $d=500$ nm measurements scales with the inverse of the respective transmission spot surfaces. This means that the totally transmitted optical power is constant.

In contrast to the NSOM tip measurements, no reflection is observed for the FIB-cut tips. This difference originates from the actual shape of the tip apex and not from the different metals used for tip coating. At 808 nm wavelength the gold and aluminum reflectance is $R^{Au}=0.976$ and $R^{Al}=0.867$, respectively. The back-reflection of the gold coated FIB-cut tips should thus even be slightly more intense than for the Al-coated NSOM tips. However, the NSOM tips show an irregular surface with bended edges, whereas the surface of the FIB-cut tips is flat with sharp edges. Therefore, efficient reflection by the FIB-cut tips would require very good parallel alignment of the fiber tips. Moreover reflection could only occur for transverse distances below the actual tip apex size.

The emission of the FIB-cut fiber tips is calculated using a straightforward electromagnetic model. This model neglects the influence of the reception tip on the field distribution, but allows to reproduce the main experimental observations. The emitting fiber tip is approximated by an apex of elliptical shape. The optical intensity inside the apex is supposed to be uniform and is represented by a homogeneous distribution of (typically $m=1164$) coherent and orthogonal electric $\bold p$ and magnetic $\bold m$ dipole pairs calculated from the incident electromagnetic field ($\bold E^i,\bold H^i$): 

\begin{equation}
	\begin{array}{l}
	\displaystyle \bold p=\frac{i}{2\pi c}\hat{\bold k}\times\bold H^1\\~\\

	\displaystyle \bold m=-\frac{i}{2\pi c}\hat{\bold k}\times\bold E^1	
	\end{array}
\end{equation}
with $\hat{\bold k}=\hat{\bold E}\times\hat{\bold H}$ the normalized wavevector and $c$ the speed of light. The emitted electrical field of one single dipole pair is given by \cite{jac98}:

\begin{widetext}
	\begin{equation}\label{eq.jack}
		\bold E(r)  =  \dfrac{1}{4\pi\epsilon_0}\left\{\left(\hat{\bold r}\times\bold p\right)\times\hat{\bold r}\dfrac{k^2}{r}+  \left[ 3\hat{\bold r}\left(\hat{\bold r}\cdot\bold p\right)-\bold p\right]\left(\dfrac{1}{r^3}-\dfrac{ik}{r^2}\right) -\dfrac{1}{Z_0}\left(\hat{\bold r}\times\bold m\right)k^2\left(\dfrac{1}{r}-\dfrac{1}{ikr^2}\right)\right\}e^{i kr}
	\end{equation}
\end{widetext}

with $\epsilon_0$ the vacuum permittivity, $Z_0$ the free space impedance, $k=2\pi/\lambda$ the free space wavevector, and $\bold r$ the distance vector from the dipole. Eq. \ref{eq.jack} takes all dipole emission terms into account, i.e. near- and far-field contributions. The optical intensity distribution is  obtained by summing the contributions over all dipole pairs and squaring the electrical field.

\begin{figure}[htb]
	\centering
	\includegraphics[width=7.5cm]{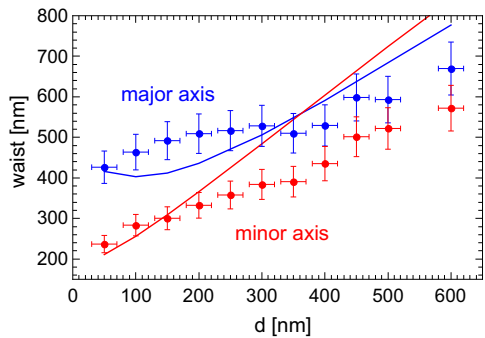}
	\caption{Major and minor axis transmission waists of the elliptical transmission spot (see Fig. \ref{fig.au}) as a function of the tip distance $d$ (points : corrected experimental data with error bars, straight lines: theory)}
	\label{fig.ecc}
\end{figure}

The measured transmission plots corresponds to the intensity emission of one tip captured by the second fiber tip. The finite apex of the second tip results in an enlargement of the transmission spot and has to be taken into account. In the present case of two different fiber tips the size correction is less straightforward than for two identical tips. Our results for the Al-coated tips show, however, that the minimal waist is of the same order as the tip apex. For correction, the triangular fiber tip is thus approximated by a circular tip of equal surface. Consequently, the calculated emission intensity plots in Fig. \ref{fig.au} are convoluted by a two-dimensional Gaussian function with a waist of 200 nm. 

The agreement of the experimental and theoretical transmission maps is satisfactory and the elliptical and nearly round shapes at respectively small and far distances are well reproduced. The corrected major and minor axis waists ($\tilde w_{a,b}$) of the elliptical transmission spots are represented on Fig. \ref{fig.ecc}. Here the experimental waists are corrected for the 200 nm apex size of the second fiber tip. The agreement of observed and calculated values is good for small tip distances and the main features are well reproduced. However, for larger tip distances the calculated values diverge from the experiment. This is mainly due to very low experimental signal levels and limits of the approximations concerning the simplified apex shapes. Also, it would be interesting to quantify the influence on the field distribution of the receiving tip, which will require full electromagnetic calculations. This is left for future work.

The minimal corrected emission waists of the elliptical fiber tip are $\tilde w^{min}_a=428$ nm and $\tilde w_b^{min}=243$ nm, slightly lower than the actual aperture of the elliptical fiber. At small distances light is transmitted by the optical near field. Thus the shape of the two tip apexes determines the shape of the observed intensity spot. For larger distances the light is transmitted by the optical far field, which cannot resolve the fiber tip shapes smaller than the diffraction limit. Thus, the obtained image corresponds to a point-like optical emitter \cite{DCH11}. This was experimentally observed by the decreasing difference between the major and minor axis waists. The numerical results even show an inversion of the ellipse major and minor axis at 365 nm, slightly below half the wavelength. 

\section{Conclusions}

In conclusion, the transmission and reflection properties of two kinds of metal-coated optical fiber tip pairs were experimentally studied. Clear evidence of optical near-field coupling between the two tips of a pair was described by sub-wavelength transverse transmission spots. This point was confirmed by the resolution of the sub-wavelength fiber tip shape. We believe that these results are of interest for the future application in optical nano-tweezers of this kind of metal-coated optical fiber tip pairs.

\section*{Acknowledgment}
Funding for this project was provided by a grant from la R\'{e}gion Rh\^{o}nes-Alpes and by the French National Research Agency in the framework of the FiPlaNT project (ANR-12-BS10-002). Helpful discussions with A. Drezet are gratefully acknowledged.


\end{document}